# Ultra-Fast Ferrimagnetic All Spin Logic Device

Yue Zhang[1,2]*, Zhizhong Zhang[1,2], Zhenyi Zheng[1,2], Jiang Nan[1,2], Guanda Wang[1,2], Kun Zhang[1,2], Yu Wang[1,2], Wei Yun[1,2], Youguang Zhang[1,2], and Weisheng Zhao[1,2]*

[1]Fert Beijing Institute, BDBC, Univ. Beihang, 100191, Beijing, China

[2]School of Electrical and Information Engineering, Univ. Beihang, 100191, Beijing, China

{yz, weisheng.zhao}@buaa.edu.cn

*Abstract*— All spin logic device (ASLD) blazes an alternative path for realizing ultra-low power computing in the Post-Moore era. However, initial device structure relying on ferromagnetic input/output and spin transfer torque (STT) driven magnetization switching degrades its performance and even hinders its realization. In this paper, we propose an ASLD based on rare-earth (RE)-transition-metal (TM) ferrimagnetic alloy that can achieve an ultra-high frequency up to terahertz. The spin orbit torque (SOT) induced fast precession near the spin angular momentum compensated point is investigated through the macrospin model. Combining the non-local spin current diffusing from the input to the output, a deterministic picosecond switching can be realized without any external magnetic field. Our results show that ASLD has the potential to exceed the performance of mainstream computing.

The new era of information raises more strict demands for electronic applications. Lower power, faster and denser are the technology development trends and also imply the bottlenecks to continue the Moore's law. Spintronics is widely regarded one of the most promising solution to overcome these bottlenecks, since it utilizes the spin degree of freedom of electrons instead of charge [1-2]. Besides the memory and storage, computing is becoming an alternative orientation of spintronics nowadays [3]. There were already various designs and prototypes of logic-in-memory circuits hybridizing spintronics and CMOS [4-5]. However, the frequent charge-spin conversions block their performance progress. The appearance of all spin logic device (ASLD) (see Fig. 1a) broadens the research horizon as it can implement the logic operations completely based on spin current [6-8]. As there is rarely charge current involving in the information processing, ultra-low power consumption can be expected. However, the realization of ASLD still confronts certain issues in terms of material, mechanism and structure. Among them, the magnetization switching of output magnet induced by a non-local spin current diffusing from the input is the most crucial one. As the mainstream concept relies on ferromagnetic materials (see Fig. 1a) and spin transfer torque (STT) effect, the spin current threshold and the switching delay are hardly reduced. Therefore, the power consumption dissipated by the spin injection will be large and the logic operation speed can also be influenced greatly [9-10].

Recently, rare-earth (RE)-transition-metal (TM) ferrimagnetic alloys, in which the RE sublattice is antiferromagnetically (AF) coupled to the TM sublattice, have been intensively investigated (see Fig. 1b) [11-13]. Compared with ferromagnets, ferrimagnets exhibit much faster dynamics. Moreover, ferrimagnets can avoid the magnetic field immunity of antiferromagnets, which can overcome the measurement and detection difficulties for practical applications [14-16]. The ultra-fast magnetic dynamics near the angular momentum compensation point in the ferrimagnetic materials is indeed ideal for building ASLD, since the computing puts more emphasis on the operational speed, compared with the memory devices [17].

In this paper, we propose an ultra-fast ASLD based on ferrimagnets (named FIM-ASLD), in which RE-TM alloy is used to construct input/output magnets (see Fig. 1b). The magnetization precession driven by spin orbit torque (SOT) is studied through the macrospin approach. The deterministic switching without external magnetic field can then be realized according to the spin direction of the non-local spin current. Through changing the sublattice concentration of the ferrimagnetic alloy, the precession frequency can be adjusted, which can reach as high as terahertz. The proposed FIM-ASLD relaxes the pressure of the spin current based devices on the spin injection and spin transfer. Due to the several orders higher frequency than the traditional ones, the reduction of power consumption can confidently be expected.

The schematic of FIM-ASLD is demonstrated in Fig. 1b. Compared with the initially proposed structure, the input/output magnets are changed from the ferromagnetic materials to RE-TM ferrimagnetic alloy. In the following studies and analyses, we apply $Tb_xCo_{1-x}$ as the ferromagnetic alloy, which can also provide strong bulk perpendicular magnetic anisotropy (PMA), unlike the interfacial PMA existing in the ferromagnetic material systems (e.g. Ta/CoFeB/MgO) [18-19]. Because of the exchange interaction between f and d electrons, the momentums of Tb sublattice and Co sublattice are aligned antiparallel. In the ferrimagnetic alloy systems, there are two important compensation points. At the magnetization compensation point, the magnetizations of two sublattices cancel each other, i.e. $M_{eff} = M_{RE} - M_{TM} = 0$. Meanwhile, the angular momentum vanishes at the angular momentum compensation point, i.e. $\frac{M_{eff}}{\gamma_{eff}} = \frac{M_{RE}}{\gamma_{RE}} - \frac{M_{TM}}{\gamma_{TM}} = 0$. These two points have different values because the gyromagnetic ratios of each sublattice are different, i.e. $\gamma_{RE} \neq \gamma_{TM}$, and they can be controlled by the sublattice concentration and temperature [11]. Another modification is that a heavy metal layer has been added to inject a spin current in the output magnets through SOT effect. As the spin angular momentum of this spin current is perpendicular to that of the ferrimagnetic magnet, the spin torque between them can drive the magnet to implement fast magnetization precession. During the logic operation, when a voltage is applied on the input magnet, spin accumulation occurs under the interface of the input magnet and the non-magnetic channel. A non-local spin current will be yielded by the spin diffusion in the channel. According to the angular momentum direction of this spin current, which is then absorbed by the output, the deterministic magnetization switching can be achieved. Therefore, through controlling accurately the voltages applied on the heavy metal layer and the input, a fast and deterministic information transfer can be realized by the cooperation of two spin currents.

To validate the functionality of the FIM-ASLD and characterize its performance, we carried out marcospin modeling to investigate the magnetization dynamics of the magnets by solving Landau–Lifshitz–Gilbert (LLG) equation with the STT and SOT terms.

$$\frac{d\mathbf{M}_{RE(TM)}}{dt} = -\gamma_{RE(TM)}\big(\mathbf{M}_{RE(TM)} \times \mathbf{H}_{effRE(TM)}\big) + \frac{\alpha_{RE(TM)}}{M_{RE(TM)}}\bigg(\mathbf{M}_{RE(TM)} \times \frac{d\mathbf{M}_{RE(TM)}}{dt}\bigg) - \gamma_{RE(TM)}\frac{J_{sot}\hbar}{2e\mu_0 M_{sRE(TM)}t}\big(\mathbf{M}_{RE(TM)} \times \boldsymbol{\sigma}_{SOT} \times \mathbf{M}_{RE(TM)}\big) + \eta\frac{\mu_B J_s}{e}\mathbf{M}_{RE(TM)} \times \big(\mathbf{M}_{RE(TM)} \times \boldsymbol{\sigma}_S\big) \qquad (1)$$

where the $\mathbf{M}_{RE}$ and $\mathbf{M}_{TM}$ are the magnetization vectors of two RE-TM sublattices, i.e. Co and Tb respectively, $\boldsymbol{\sigma}_{SOT}$ is the spin moment vector of the spin current generated by SOT，$\boldsymbol{\sigma}_S$ is the spin moment vector of the spin current absorbed by output magnet, which induces STT on its magnetization. The other parameters and their default values are listed in Table. I.

We can then describe the net magnetization of the ferrimagnetic alloy as follows.

$$\frac{d\boldsymbol{M}_{eff}}{dt} = -\gamma_{eff}(\boldsymbol{M}_{eff} \times \boldsymbol{H}_{eff}) + \frac{\alpha_{eff}}{M_{eff}}\left(\boldsymbol{M}_{eff} \times \frac{d\boldsymbol{M}_{eff}}{dt}\right) - \gamma_{eff}\frac{J_{sot}\hbar}{2e\mu_0 M_2 t}(\boldsymbol{M}_{eff} \times \boldsymbol{\sigma}_{SOT} \times \boldsymbol{M}_{eff}) +$$

$$\eta\frac{\mu_B J_s}{e}\boldsymbol{M}_{eff} \times (\boldsymbol{M}_{eff} \times \boldsymbol{\sigma}_S) \quad (2)$$

where $\gamma_{eff}$ and $\alpha_{eff}$ can be expressed as

$$\gamma_{eff} = \frac{M_{RE} - M_{TM}}{\frac{M_{RE}}{\gamma_{RE}} - \frac{M_{TM}}{\gamma_{TM}}} \quad (3)$$

$$\alpha_{eff} = \frac{\alpha_{RE}\frac{M_{RE}}{\gamma_{RE}} + \alpha_{TM}\frac{M_{TM}}{\gamma_{TM}}}{\frac{M_{RE}}{\gamma_{RE}} - \frac{M_{TM}}{\gamma_{TM}}} \quad (4)$$

First, we investigate the case without the spin current injected by the heavy metal layer. In this case, the output magnet only absorbs the spin current diffusing from the input magnet, which is like the spin transport mechanism in the initial ASLD concept based on ferromagnetic materials. We thus set the SOT induced spin current $J_{sot}$ to 0 in the model above. As shown in Fig. 2a, $\boldsymbol{M}_{TM}$ and $\boldsymbol{M}_{RE}$ are always anti-parallel in the dynamic process due to the strong antiferromagnetic coupling. The precession trajectory of the major magnetization $\boldsymbol{M}_{TM}$ is similar to that of magnetization under STT effect. This structure and configuration cannot exhibit obvious performance advantages compared with the previous concepts.

According to Eq. 3, when $\frac{M_{RE}}{\gamma_{RE}} - \frac{M_{TM}}{\gamma_{TM}} = 0$, i.e. at the angular momentum compensation point, $\gamma_{eff}$ will be divergent. At the same time, if we apply a spin current perpendicular to the magnetization vector of the Co and Tb sublattices, i.e. along the hard axis of the material, an ultra-fast precession under the generated demagnetization field can thus be stimulated. In our proposed structure, a heavy metal layer is introduced to inject the required spin current through SOT effect. From the point of view of energy economization and integration, this spin current injection is also superior to magnetic field that has always been used to induce magnetization precession in ferrimagnetic materials in the previous literatures. As shown in Fig. 2b, with the effort of the spin current aligned in y-axis, the projection of the two sublattices magnetizations in z-axis exhibits the precession behavior. Meanwhile, the sublattice concentration has an obvious impact on the precession frequency. When the concentration is close to the angular moment compensated point, the frequency will become extremely high. As shown in Fig. 2c, by tuning the concentration ratio x in $Tb_xCo_{1-x}$ from 0.15 to 0.165, the precession frequency can be increased from 53.08 GHz to 0.21 THz. This precession provides us the potential to realize terahertz information processing.

Beyond the ultra-fast precession, the deterministic magnetization switching can be achieved by applying a voltage on the input. This voltage can implement spin accumulation and a spin current can diffuse from the input to the output. As the final state of the magnetization always trends to be the one making the system with the energy minimum, this switching process can be investigated by calculating the energy of the system, which can be written as

$$\mathrm{E} = -K_{eff}\int \sin^2\theta\, dr - \frac{\mu_0}{2}\int H_d \cdot M_{eff}\, dr \quad (5)$$

where $\theta$ is the angle between the magnetization and the anisotropy easy axis, $K_{eff}$ is the effective anisotropy constant, $H_d$ is the demagnetization field. As demonstrated in Fig. 3a, the energetically favored direction of the major magnetization vector is at an unstable state in the magnetization precession, and the magnetization can be halted in a random state. If there is another spin current, aligned either parallel or antiparallel to the magnetization of the output magnet, a deterministic lowest energy point will occur at one of two stable states of the system (see Fig. 3b). Correspondingly, the final magnetization direction can be determined. Compared with the relatively slow switching delay of ferromagnet that can hardly be reduced below 0.1 ns, the switching delay of ferrimagnet can reach as low as some picosecond (see Fig. 4). Here, we discuss the power consumption of FIM-ASLD qualitatively. Compared with the initial concept, an additional current required for inducing magnetization precession will increase somewhat the energy consumption. However, this degradation can be alleviated or neglected based on the following considerations. First, the switching delay or information transfer time can be greatly reduced by adding this current, normally several orders. The energy consumption can certainly be reduced due to this significant frequency improvement by applying extremely short pulses. Second, the additional current can be shared by numbers of the bits (see Fig. 1c). This sharing can be beneficial to the energy economization of future massive spin computing system [20].

We propose an ASLD based on ferrimagnetic alloy. By combining two orthogonal spin currents generated by spin diffusion and SOT, ultra-fast (up to THz) deterministic magnetization switching can be realized. Macrospin modeling has been applied to validate its functionality and analyses its performance. Compared with the initial structure based on ferromagnetic materials, the proposed FIM-ASLD can effectively increase frequency, reduce energy consumption as well as relax the non-local spin injection pressure. This work proves that ferrimagnetic spintronics is ideal for realizing ultra-fast and ultra-low-power computing.

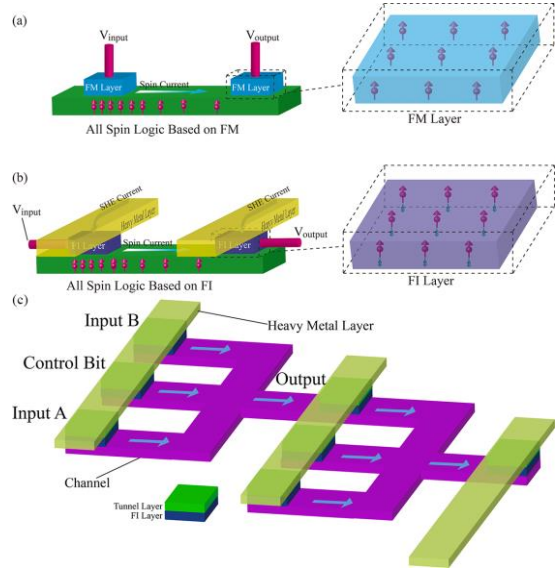

FIG. 1. (a) Schematic diagram of initial ASLD structure relying on ferromagnetic materials. (b) Schematic diagram of ASLD based on the ferrimagnets, where a heavy mental layer is added to generate SOT spin current. (c) Cascade of ASLDs for implementing complex logic functions.

TABLE I

PARAMETERS AND VARIABLES PRESENT IN THE FITTING FUNCTIONS

| Parameter | Description | Default Value |
|---|---|---|
| $M_{sRE}$ | RE saturation magnetization | 1.40x $10^6$ A/m |
| $M_{sTM}$ | TM saturation magnetization | 0.95 x $10^6$ A/m |
| $\gamma_{RE}$ | RE gyromagnetic ratio | -1.9347 x $10^{11}$ $T^{-1}s^{-1}$ |
| $\gamma_{TM}$ | TM gyromagnetic ratio | 1.3191 x $10^{11}$ $T^{-1}s^{-1}$ |
| $\alpha_{RE}$ | RE gilbert damping | 0.02 |
| $\alpha_{TM}$ | TM gilbert damping | 0.019 |
| $\mu_B$ | Bohr magneton | 9.274 x $10^{24}$ J*$T^{-1}$ |
| $J_{sot}$ | SOT spin current density | 8 x $10^{11}$ A/$m^2$ |
| $J_s$ | Diffused spin current density | 8 x $10^{10}$ A/$m^2$ |
| η | Spin polarization | 0.5 |
| e | Elementary charge | 1.6 x $10^{-19}$ C |
| t | Free layer thickness | 1 nm |

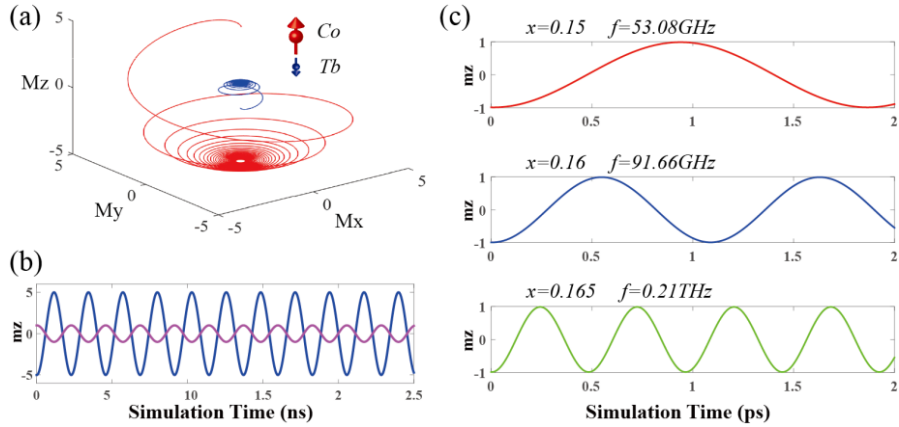

FIG. 2 (a) Magnetization trajectory of two sublattices induced by the spin current diffused from the input. (b) Magnetization precession of two sublattices. (c) The precession frequency continues to increase near the spin angular momentum compensated point by changing concentration ratio.

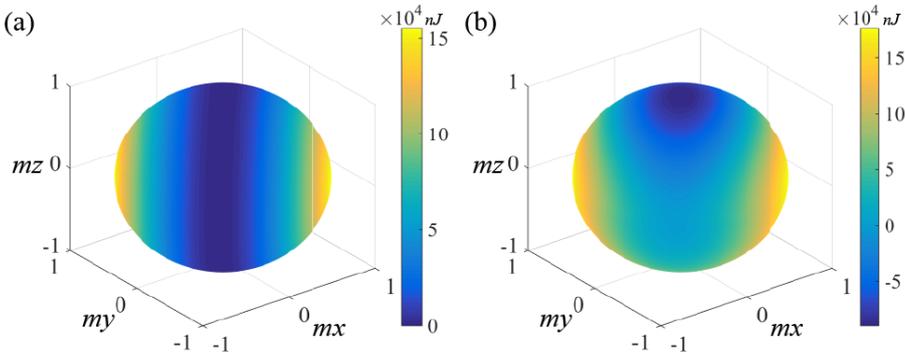

FIG. 3 Diagram of the system energy distribution depended on the direction of the major magnetization vector. (a) Without applying the input voltage. (b) With applying the input voltage.

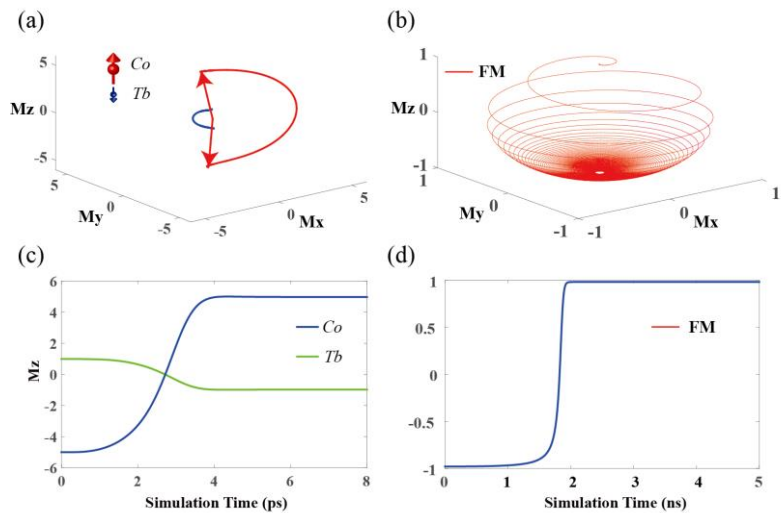

FIG. 4 Performance comparisons between the FIM-ASLD and ferromagnetic ASLD. (a) Trajectory of the magnetization vector. (b) Switching speed.